\documentclass{pasj01}
\input{epsf}

\Received{$\langle$reception date$\rangle$}
\Accepted{$\langle$acception date$\rangle$}
\Published{$\langle$publication date$\rangle$}

\begin{document}

\title{Characteristics of mid-infrared PAH emission from star-forming galaxies selected at 250 $\mu$m in the North Ecliptic Pole (NEP) field }
\author{Seong Jin Kim$^{1,2}$}
\author{Woong-Seob Jeong$^{2}$}  
\author{Tomotsugu Goto$^{1}$}
\author{Hyung Mok Lee$^{2}$}
\author{Hyunjin Shim$^{3}$} 
\author{Chris Pearson$^{4,5}$}  
\author{Myungshin Im$^{6}$} 
\author{Hideo Matsuhara$^{7}$}  
\author{Hyunjong Seo$^{2}$} 
\author{Tetsuya Hashimoto$^{1}$}
\author{Minjin Kim$^{2}$}
\author{Chia-Ying Chiang$^{1}$}
\author{Laia Barrufet De Soto$^{4,5}$}
\author{Maria del Carmen Campos Varillas$^{4}$}

\altaffiltext{1}{National Tsing Hua University No. 101, Section 2, Kuang-Fu Road, Hsinchu, Taiwan 30013}
\altaffiltext{2}{Korea Astronomical Space Science Institute, 61-1, Whaam-dong, Yuseong-gu, Deajeon 305-348, Korea}
\altaffiltext{3}{Department of Earth Science Education, Kyungpook National University, Deagu 702-701, Republic of Korea}
\altaffiltext{4}{RAL Space, Rutherford Appleton Laboratory, Chilton, Didcot, Oxfordshire OX11 0QX, UK}
\altaffiltext{5}{The Open University, Milton Keynes, MK7 6AA, UK}
\altaffiltext{6}{Astronomy Program, FPRD, Department of Physics and Astronomy, SNU, Kwanak-Gu, Seoul 151-742,  Korea}
\altaffiltext{7}{Institute of Space and Astronautical Science, Japan Aerospace Exploration Agency, Yoshinodai 3-1-1, Sagamihara, Kanagawa 229 8510}

\email{seongini@gmail.com}


\maketitle

\begin{abstract}
 Evolutionary properties of infrared (IR) luminous galaxies are important keys to understand 
 dust-obscured star formation history and galaxy evolution. Based on the near- to mid-IR 
 imaging with 9 continuous filters of AKARI space telescope, we present the characteristics of 
 dusty star-forming (SF) galalxies  showing  polycyclic aromatic hydrocarbon (PAH) features observed
 by the North Ecliptic Pole (NEP) wide field survey of AKARI and \textit{Herschel}.  All the 
 sample galaxies from the AKARI/NEP-Wide data are selected based both on the Herschel/SPIRE 
 250 $\mu$m detection and optical spectroscopic redshift data. The physical modelling of spectral 
 energy distribution (SED)  using all available data points from u$^*$ to sub-mm 500 $\mu$m band, 
 including WISE and PACS data where available, takes unique advantages of the continuous near- to 
 mid-IR coverage, reliable constraint on far-IR peak, spectroscopically determined accurate 
 redshifts, as well as energy balance principle by \texttt{MAGPHYS}. This enables us to derive 
 physically meaningful and accurate total infrared luminosity and 8 $\mu$m (or PAH) luminosity 
 consistently.  Our sample galaxies are in the redshift range z $<$1, and majority of them appear
 to be normal SF/spiral populations  showing  PAH features near the 8 $\mu$m.   These SF galaxies 
 showing PAHs in the mid-IR include various types from quiescent to starbursts.   Some of our 
 sample show shortage of 8 $\mu$m luminosity compared to the total IR luminosity and this PAH 
 deficit gets severe in more luminous IR galaxies, suggesting PAH molecules in these galaxies 
 destroyed by strong radiation field from SF region or a large amount of cold dust in ISM.  
 The specific SFR of our sample  shows mass dependent time evolution which is consistent with 
 downsizing evolutionary pattern.

\space
  
\end{abstract}

\section{Introduction}
Infrared (IR) luminous population of galaxies are key objects to fully understand the evolution 
of galaxies and cosmic star formation history. Among various evolutionary properties of IR galaxies, 
total infrared luminosity (L$_{IR}$) is an important and representative quantity that gives overall 
impression of a galaxy at a glance. The relation between L$_{IR}$ and other galaxy parameters 
(e.g., star formation rate, dust mass, mid-IR luminosities, stellar mass, etc.) is  always an 
interesting issue and  a matter of great concern in modern astronomy.  Especially, the relation with 
mid-IR (MIR) properties such as 8$\mu$m luminosity ($L_{8{\mu}m}$), representing brightness of 
polycyclic aromatic hydrocarbon (PAH) features in the MIR, is one of the interesting issues because 
the MIR spectra of most normal galaxies are dominated by  PAH emissions at  6.2, 7.7, 8.6, 11.3 and 
12.7 $\mu$m.  They are large carbonated molecules believed to be ubiquotous in the interstellar 
medium (ISM) and known to glow mostly in photo-dissociation regions (PDRs), at the interface between 
the ionized and molecular gas (the outskirts of HII regions) by cooling down through numerous 
rotational and vibrational modes, thus producing a group of bright and broad emissions in the MIR.  
It is needless to say that accurate measurement of these quanties ($L_{IR}$ and rest-frame $L_{8{\mu}m}$) 
is essential in order to reveal the mysterious connection between  overall strength of PAHs around 
8$\mu$m and the physical condition of star-forming (SF) activities (Smith et al. 2007; Elbaz et al. 
2011; Schreiber et al. 2018),  or to get any closer to one small clue at least.   However,  
convincing determination of both $L_{IR}$ and $L_{8{\mu}m}$  from various unbiased SF galaxy 
(SFG) sample is normally limited due to insufficient photometric data spanning a wide range of 
multi-wavelength bands consistently covering large sky field and lack of accurate redshifts 
information.

The past decade of IR astronomy with space mission such as AKARI (Murakami et al. 2007) and Spitzer  allowed us to have extensive IR wavelength coverage  and to achieve the meaningful and more reliable constraints on the IR emission of galaxies. Especially, continuous wavelength coverage from 2 to 24 $\mu$m by AKARI is unique for MIR luminosity (or PAH luminosity) measurement  that we can not get elsewhere.  Also \textit{Herschel} Space Observatory (Pilbratt et al. 2010) has extended wavelength range and  offered valuable far-IR (FIR) data with the Photodetector Array Camera and Spectrometer (PACS, Poglitsch et al. 2010) and revolutionary sub-millimeter (sub-mm; SMM) data with the Spectral and Photometric Imaging Receiver (SPIRE, Griffin et al. 2010), giving information on the FIR peak that dominates the spectral energy distributions (SEDs) of IR luminous populations.  With expanding interests in dust residing in IR galaxies, all the wavelengths are getting crucial in terms of more physically motivated SED modelling and for the better understanding of formation and evolution of dusty star forming galaxies.  In addition, the detection  at the \textit{Herschel}/SPIRE bands provides an important basis to select various types of IR luminous populations of galaxies.

 The North Ecliptic Pole (NEP) region surveyed by \textit{Herschel}'s SPIRE instrument is one of the legacy fields observed by \textit{AKARI} (Matsuhara et al. 2006; Murakami et al. 2007),  composed of two-tier survey programmes; the NEP-Wide (NEPW) survey (covering $\sim$ 5.4 deg$^2$,  Lee et al. 2009; Kim et al. 2012) and the NEP-Deep (on a smaller $\sim$ 0.67 deg$^2$ with higher integration time) survey (Wada et al. 2008; Takagi et al. 2012; Murata et al. 2013).  
 One of the interesting issues on the NEP field was derivation of mid-IR (MIR) luminosity functions (LFs, Goto et al. 2010, 2015; Kim et al. 2015) of star forming galaxies over the near and far universe,  in the context of the $L_{8{\mu}m}$  connection to the SF activities.  Thanks to the  \textit{Herschel}/SPIRE survey on the NEP, we can perform the analysis of the sub-mm (SMM) counterparts of those dusty SF galaxies. With the spectroscopic data on this field (Shim et al. 2013), we can simultaneously use the information of optical emission lines, detailed MIR features as well as dust continuum with FIR/SMM peaks,  produced by dust and gas heated by stars, all of which are crucial to understand large picture of the dusty star-forming activities. 

  
 In this work, our aim is to investigate evolutionary properties of local ($z < 1$) dusty star-forming galaxies having MIR PAH features. Using AKARI data, more accurate derivation of 8$\mu$m luminosity (dominated by PAH features) and comparison with total infrared luminosity are critical procedures in this work. In the sec. 2, we describe data and properties of sample.  We show how we carried out modelling to obtain physical quantities in sec. 3. In the next section, we present the results and discussions on their properties in diversified perspectives - luminosity, mass, SFR and so on. We summarize this work in sec 5.

\section{Data and Sample }

\subsection{ Multiwavelength Data Based on the AKARI and \textit{Herschel}/SPIRE}

This work used the \textit{Herschel}/SPIRE survey of the AKARI's North Ecliptic Pole (NEP) field.  The AKARI, an IR space telescope by JAXA/ISAS has left a legacy, valuable IR data obtained from a successful mission towards the NEP, covering $\sim$ 5.4 deg$^2$ wide circular area centered on the NEP (thereby NEP-Wide or NEPW). The spectral range spanning 2 -- 25 $\mu$m  continuously covered with nine photometric bands of infrared camera (IRC, Onaka et al. 2007)  provide us with unique advantage by filling the MIR gap between the \textit{Spitzer} IRAC and MIPS, making these surveys distinct from any others.  The Herschel carried out the 9 deg$^2$ survey over this field as an open time 2 program (OT2, in 2012), and completely covered the entire NEP-Wide field, with the SPIRE instruments at the 250 (PSW), 350 (PMW), and 500 (PLW) $\mu$m (achieving  9.0, 7.5, and 10.8 mJy sensitivities at each band).
Source extraction was carried out on the 250 $\mu$m map and approximately $\sim$ 4800 sources were catalogued with photometry in all three SPIRE bands. The detailed description of the data reduction,  photometry, and source properties, such as galaxy colors and source counts, can be found  in Pearson et al. (\textit{in prep.}) .

\begin{figure}
  \includegraphics[width=\columnwidth]{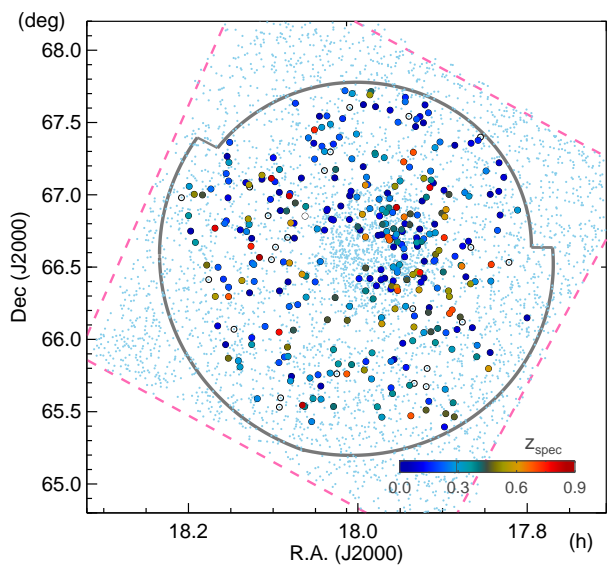}
    \caption{The North Ecliptic Pole (NEP) field surveyed by AKARI/IRC and \textit{Herschel}/SPIRE.  Outermost dashed rhombus (in magenta) indicates the area observed by the SPIRE and many small dots (in cyan) inside this large rhombus show the spatial distribution of all the 250 $\mu$m sources detected by SPIRE (PSW; at 250 $\mu$m).   A circular area (in gray) inside the rhombus is AKARI's NEP-Wide field. The filled circles in various colors show the distribution of the sample galaxies used in this work, i.e., the sample with spectroscopic redshifts (spec-$z$ $< 1$) as shown by colorbar (the black hollow circles represent sources at `$z > 1$',  which are active galactic nucleus (AGN) types, not used in this work).  } 
    \label{fig:example_figure}
\end{figure}

Fig. 1 presents a map showing the two survey areas towards the NEP by AKARI and Herschel. An outermost rhombus marked with dashed magenta line indicates the area surveyed by the Herschel/SPIRE. The circular region described by solid gray line represents the NEP-Wide (NEPW) Survey area observed by AKARI/IRC. The filled circles in colors show the distribution of our sample used in this work.

One of strong advantages on the NEP region is abundant ancillary data such as high-quality optical/NIR data obtained from ground-based observations (Hwang et al. 2007; Jeon et al. 2010; Jeon et al. 2014). Various X-ray, UV, sub-mm, and radio data over a limited region (Kollgaard et al. 1994; Lacy et al. 1995; Sedgwick et al. 2009; White et al. 2010; Krumpe et al. 2015; Geach et al. 2017) are also available.  Especially, spectroscopic follow-up survey were carried out, mainly targetting  MIR ($11\mu$m) selected SF galaxies  over the entire NEPW field (Shim et al. 2013), thereafter various emission lines (H$\alpha$, [NII], [OIII], [SII], etc.) are identified and redshifts are measured for $\sim$ 1,700 targets. This spectroscopic data has been combined with  photometric  data (from $u^{*}$ to 24 $\mu$m bands; Kim et al. 2015) to derive local MIR LFs.


\subsection{Properties of sample}

Our sample is AKARI-NEPW sources and selected based on the SPIRE 250 $\mu$m detection as well as  the spectroscopic redshift (spec-z).  The redshift data is from the optical spectroscopic survey carried out over the NEPW field  with MMT/Hectospec and WIYN/Hydra (Shim et al. 2013). The target selection for this spectroscopic survey was primarily based on MIR fluxes  ($S11 <18.5$ mag or $f_{S11} > 150 \mu$Jy).  This spectroscopic data combined with AKARI-NEPW band-merged catalogue (Kim et al. 2012) was matched to the SPIRE data (Pearson et al. \textit{in prep.}). This procedure gave us $\sim$ 300 sample; 250 galaxies at $0.0 < z < 0.9$ (with median at 0.3), 32 AGN types at $0.0 < z < 3.4$, and the others (a few galactic stars and unidentified sources). When we carried out matching, we checked all the images from the optical to the SPIRE bands.  Sources having multiple counterparts were also excluded in this procedure. We excluded AGN types and take SFGs only.   The number distribution of sample as a function of 250$\mu$m (PSW) flux are summarized in the fourth panel of Fig. 2. (The distribution as a fuction of redshfit is shown at the first panel of Fig. 4.)  


\begin{figure}     
  \includegraphics[width=0.495\columnwidth]{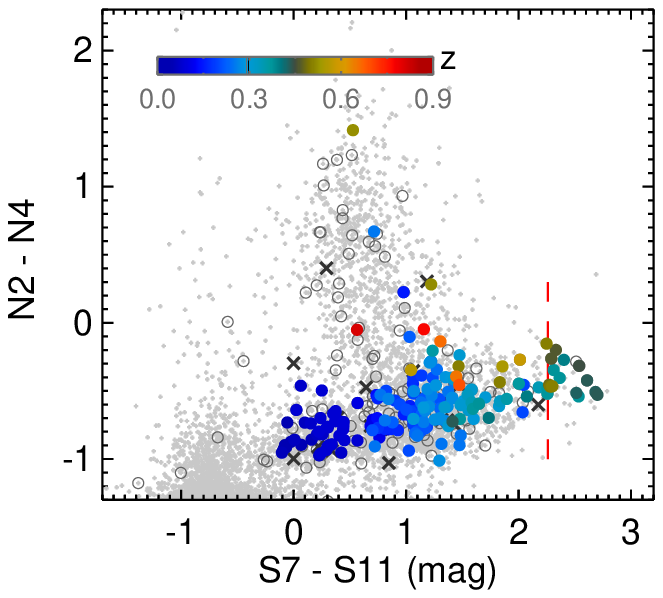}
  \includegraphics[width=0.495\columnwidth]{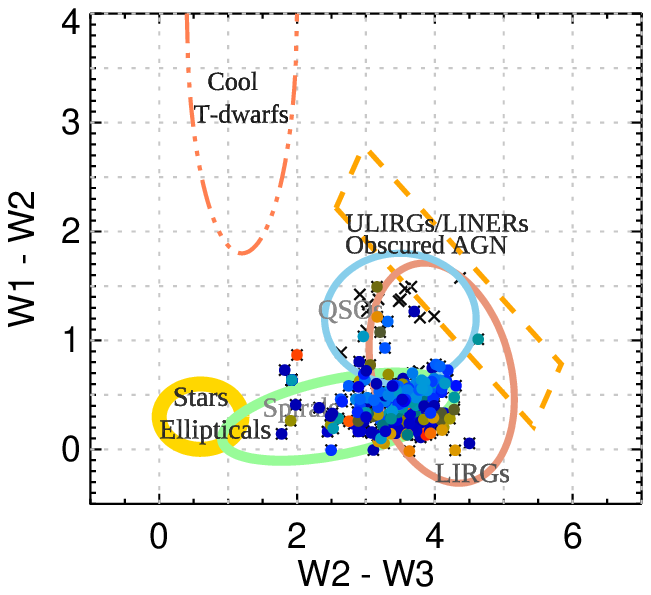}
  \includegraphics[width=0.495\columnwidth]{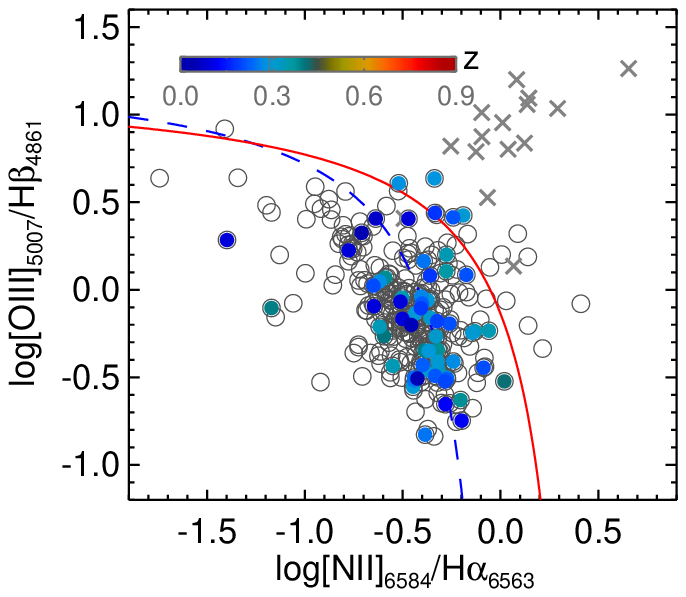}
  \includegraphics[width=0.495\columnwidth]{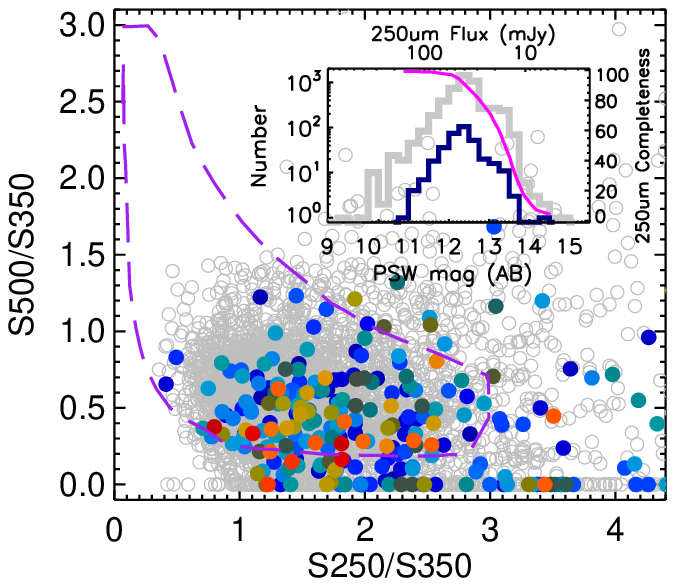}
    \caption{color-color diagrams (CC Diags) based on various photometric bands and BPT (Baldwin et al. 1981) diagram for emission lines to show the characteristics of the selected sources.  In all panels, filled circles (see color bar) indicate the SPIRE 250$\mu$m detected galaxies having spectroscopic redshift ($z < 1$). Top left: Our sample in the AKARI/IRC CC diag. Many background gray dots represent the AKARI sources from the NEPW catalogue (Kim et al. 2012).  `\textsf{O}'and `\textsf{X}' symbol are  250 $\mu$m sample with spec-$z$ ($ > 1$)   classified as `galaxy' and `AGN' types, respectively. 
    Dashed vertical line (in red) indicates the color cut for the MIR PAH selection (Takagi et al, 2010).  Top right: WISE CC diag showing that most of our sample are located in the overlapped area by sprial/starforming/LIRGs type populations.  Bottom left: BPT diagram showing diagnostics for emission lines ratios.  Background gray plots are all the spectroscopic sources from Shim et al. (2013). Filled circles with colors are our sample galaxies having 250$\mu$m detection. Red (solid) and blue (dashed) curves represent the scheme for classifying galaxies using emission-line ratios (Kewley et al. 2006).  Bottom right: 250 $\mu$m histogram and color-color diagram for Hershel/SPIRE sources. Gray histogram inside the box and open circles in the CC diag are all SPIRE sources in the SPIRE/NEP catalogue, and navy blue histogram and filled circles with various colors represent distribution of our sample. Magenta curve indicates 250 $\mu$m completeness.
     }
    \label{fig:example_figure}
\end{figure} 

  We present the optical/IR to SMM properties using color-color diagrams (CC Diags) based on various photometric bands from AKARI, WISE and SPIRE as well as Baldwin-Phlips-Terlivich (BPT) diagram (Baldwin et al, 1981) by optical emission lines ([OIII]/H$_{\beta}$ vs [NII]/H$_{\alpha}$) as shown in fig. 2.   In all these panels,  solid circles filled with colors  represent 250 $\mu$m  SPIRE sample having spec-$z$ ($< 1$, see colorbar) and photometric/spectroscopic information at each band. Top left (1st) panel shows NIR $(N2-N4)$ vs MIR $(S7-S11)$ color-color diagram based on the AKARI/IRC bands.  Many background gray dots represent the AKARI/IRC sources from the NEPW catalogue. `\textsf{O}' and `\textsf{X}' symbols (also plotted in 3rd panel) are showing 250 $\mu$m sample with higher spec-$z$ ($ > 1$, excluded in this work).
  A vertical dotted line (in red) shows a criterion for finding galaxies with prominent MIR PAH features (see Takagi et al. 2010).  Most of our sample is located on the left side of this vertical line, which means their PAH emissions are not very strong. 
  Top-right (2nd) panel is a well-known color-color diagram  based on the WISE bands (Wright et al. 2010), indicating a large fraction of our sample appears to be  spiral/starbursts  or Luminous Infrared Galaxies (LIRGs) populations. 
In bottom left (3rd) panel, the diagnostic (or BPT) diagram of the line ratios shows the classification scheme of galaxies using different excitation mechanism (Kewley et al., 2006; Shim et al. 2013) and actual location of our sample. 
Compared to all the spectroscopic sample (Shim et al. 2013), our SF galaxies having all these 4 line information, therefore plotted in the 3rd diagram with colorbar, are in the redshift range of $z < 0.4$. The comparison of these line emissions of higher-z ($>0.4$) sample is not possible.   The bottom-right panel shows SPIRE color-color diagram, in which all open circles are from the SPIRE NEPW catalogue. In the color-color space, dashed purple lines represent the averaged area occupied by various model SEDs generated based on modified black body (Amblard et al. 2010).  A small box inside this CC diag, we present histogram of sample used in this work (navy-blue line) as a function of 250$\mu$m flux. Here, a magenta curve indicates 250 $\mu$m completeness.  These spec-$z$ matched SPIRE sample are  250 $\mu$m counterparts of AKARI/MIR selected star-forming galaxies showing prominent PAH features in the mid-infrared. Also, they appear to be various type of low-z ($< 1$)  SFGs/LIRGs populations, normally showing 10 -- 100 mJy flux levels at sub-mm (250 $\mu$m) bands.

\section{Analysis: Modelling of SED  }

SED-fit is a critical methodology when we want to determine the physical properties using a wide range of multiwavelength photometric data. To derive the physical quantities of these SPIRE 250 $\mu$m detected and AKARI/MIR selected galaxies, we used all available data from the CFHT $u^{*}$ to SPIRE 500 $\mu$m bands (24 data points, at most).  We carried out the SED fit  using publicly available code \texttt {MAGPHYS} (da Cunha et al. 2008), which assumes an energy balace between absorbed stellar light and reradiated IR emission by dust. It offers largely empirical and physically motivated models to interpret the spectral energy distribution of galaxies. In that way, the SED modelling by \texttt {MAGPHYS}  can describe various galaxy SEDs from uv to 1000 $\mu$m, based on the stellar emission (including dust attenuation, see Bruzual \& Charlot, 2003; which adopted the parametrization by Chabrier (2003, PASP, 115, 763) of the single-star IMF in the Galactic disc), 
the PAH templates in the mid-IR  (Li $\&$ Draine 2001; Madden et al. 2006), and  dust emission from the stellar birth clouds (BC), $L_{d}^{BC}$  and from the ambient interstellar medium (ISM), $L_{d}^{ISM}$ (see Charlot \& Fall 2000; da Cunha et al. 2008). Determination of $L_{8{\mu}m}$ based on the SED-fitting is not a direct measurement of PAH strength, but this is best ways to get $L_{IR}$ and $L_{8{\mu}m}$ consistently  together with the other physical quantities such as star-formatio rate (SFR), stellar mass (M$_{\ast}$), for a large number of galaxy sample,  especially thanks to the continuous 9-band photometry working as a low-resolution spectra.

The original code doesn't take AGN component into the IR emission and adopts the fixed template spectrm of PAHs (Li $\&$ Draine 2001; Madden et al. 2006) because of no need to consider variations by AGNs. Also it doesn't perform photometric redshifts (phot-$z$) estimation.  These facts make it clear that using this code is so useful and adequate for our SED-fit procedure because our sample has spectroscopic redshift and type was classified as galaxy by emission-line diagnostics (Shim et al. 2013) so that we can start with simple assumption that all the sample galaxies (having accurate redshift) emit IR emission predominantly (or purely) by star-forming activities.

\begin{figure}
  \includegraphics[width=0.495\columnwidth]{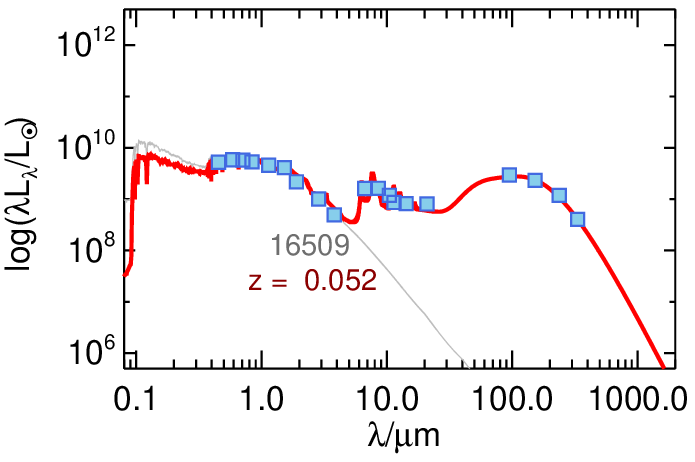}
  \includegraphics[width=0.495\columnwidth]{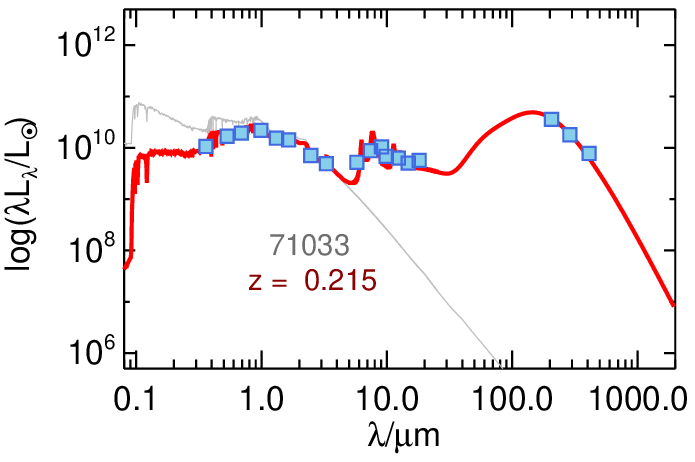}
  \includegraphics[width=0.495\columnwidth]{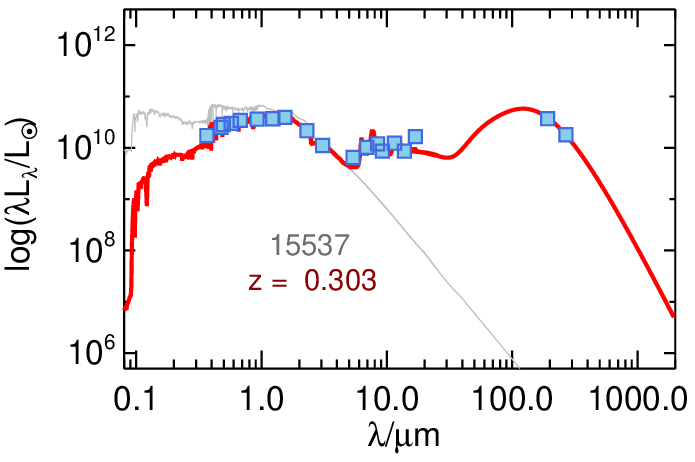}
  \includegraphics[width=0.495\columnwidth]{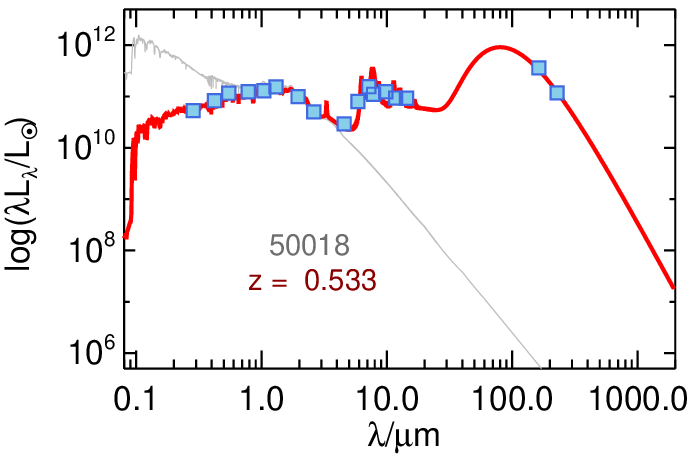}
   \caption{ Examples of SED-fit, presented in units of the rest frame wavelength and luminosity ($\lambda L_{\lambda}/L_{\odot}$). The red lines represent the best-fit  obtained from `\texttt{MAGPHYS}'. The gray lines show the unattenuated stellar spectra, and the filled squares indicate photometric data points from the CFHT, Maidanak, KPNO, AKARI, and \textit{Herschel}/SPIRE and PACS (where available).
   } 
    \label{fig:example_figure}
\end{figure}

As shown in Fig. 3, this physical modelling provides flexibility in the SED-fit throughout our photometric data points, giving insight for interpretation of SED properties (e.g., FIR peak, dust temperature ($T_{d}$), galaxy type, etc.) without any superfluous assumptions.  Optical to NIR data points sampled by various (CFHT, Maidanak, KPNO, and AKARI) observations are fitted well mostly by the sum of the obscured stellar emission plus a small fraction of NIR continuum from hot dust (small grains suffering from stochastic heating up to only a few hundreds kelvin), and MIR points are by sum of PAH emisison and underlying continuum from dusts,  while the FIR data points are desibed by a broad peak  generated by emission from dusts in thermal equilibium (warm and cold dust in BC and ISM, respectively), suggesting that simple isothermal component can hardly describe wide range of FIR-SMM SEDs of galaxies, requiring secondary (or more) component, due to the continunuous distribution of size and temperature.  This also explains one of aspects why a fixed number of some earlier template libraries can not avoid difficulties when fitting the data from uv to SMM. 

Among our sample, 25 $\%$ has PACS (Pearson et al. 2018 \textit{submitted to this issue}) counterparts and photometric data at 100 and/or 160 $\mu$m.  SED-fit including PACS data shows several $\%$  differences in terms of the total infrared luminosity ($L_{IR}$), compared to the results without them (showing no tendency to have   higher or lower $L_{IR}$,  nor  shift of FIR peak toward longer or shorter wavelength).  Existence of the WISE data doesn't make remarkable differences in the SED-fit results.


\section{Results and Discussion}

The Physical modelling by $\texttt{MAGPHYS}$ gives important physical quantities such as star-formation rate (SFR), stellar mass ($M_{\ast}$), total dust luminosity (L$_{d}^{tot} = L_{d}^{BC} +L_{d}^{ISM}$), dust mass ($M_{d}$).   Total IR luminosity ($L_{IR}$) is obtained by integrating rest-frame SED from 8 to 1000$\mu$m ($\int _{8 {\mu}m} ^{1000} L_{\lambda}d\lambda$) , and   8$\mu$m luminosity ($L_{8{\mu}m}$ defined by ${\nu}L_{\nu}$ ($8{\mu}m$) ) is derived by convolving the rest-frame SED with the Spitzer 8$\mu$m filter response.  L$_{d}^{tot}$  (sum of all dust emission)  is  almost the same as  L$_{IR}$ for each galaxy (within a few per cent), obviously showing what's the origin of galaxy IR emission.

 \begin{figure}
 \includegraphics[width=1.01\columnwidth]{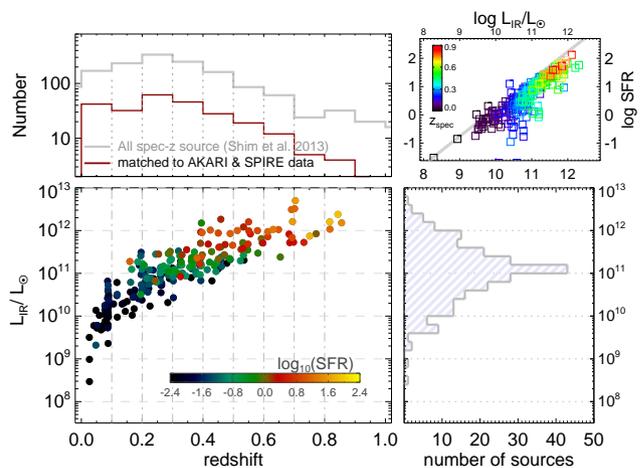}
    \caption{ The number distribution of sample as a function of redshift and total infrared luminosity ($L_{IR}$). Top panel shows the histogram of SPIRE 250 $\mu$m galaxies matched to the optical spectroscopic survey data. Bottom-left panel shows the distribution of total infrared luminosity ($L_{IR}$) for our galaxies. In this panel, the color represents  star-formation rate (SFR, in logarithmic scale) derived from the model fitting. Bottom-right panel shows the histogram for $L_{IR}$, which peaks at  $\sim 10^{11} L_{\odot}$ (the LIRG threshold).  
    } 
    \label{fig:example_figure}
\end{figure}

In Fig. 4, we present the number distribution of our  galaxies in term of  L$_{IR}$ and SFR derived from the SED-fit.  Top left panel summarizes the total number of  spectroscopic sample (gray solid line) in NEP-Wide field and those matched to AKARI and SPIRE data (dark red), showing  how the number is reduced and left with $\sim$ 250 sources, as a function of redshift.  Bottom left panel shows the distribution of $L_{IR}$ according to redshift, with the comparison of star-formation rate (SFR) derived by \texttt{MAGPHYS} (indicated by a colorbar).  The bottom right panel shows the number of sample in terms of L$_{IR}$, which peaks around $10^{11} L_{\odot}$ (the criterion of LIRGs).  More than a half ($\sim$ 55$\%$) of the sample turned out to be IR luminous populations (here, LIRGs ($49\%$) and ULIRGs ($6\%$) out of all sample). These three diagrams  partly share common axis among them. Additional plot, the top right panel, shows the relation between  L$_{IR}$ and SFR with a  colorbar of redshfit. This plot is a different version of bottom left panel.  A gray line in the background represent the Kennicutt (1988) relation to estimate SFR from L$_{IR}$, which can be used as if all IR emission comes from SF activity. \texttt{MAGPHYS} estimates SFR using the energy balance between obscured stellar light and dust re-emission, on the basis of the modeling in which the coldest dust component is not included in the stellar birth clouds and heated by old ($> 10^7$ yr) stars in the ISM, thus, not involved in actual SF activity.   That is why the SFRs estimated in this work do not have a tight linear correlation with L$_{IR}$.  This line can be regarded as an upper limit of SFR estimation from the $L_{IR}$ measurement.

\begin{figure}
 \includegraphics[width=1.0\columnwidth]{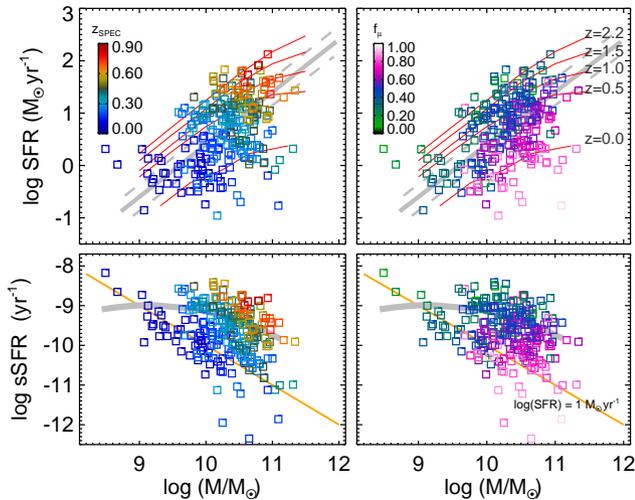}
    \caption{Distribution of our galaxies in the plane of (specific) star-formation Rate (SFR) vs  stellar mass ($M_{\ast}$) with colorbar showing  redshift (left panels) and parameter f$_{\mu}$ (right panels). In upper panels, thick gray lines in the background represent the IR galaxy main-sequence (MS) and its 1-$\sigma$ dispersion (dashed line) from Elbaz et al. (2011).  Magenta lines indicate the MS evolution from $z = 0$ to $z = 2$ (Schreiber et al. 2016). In lower panels, the sample is shown in terms of specific SFR (sSFR) vs stellar mass ($M_{\ast}$). Gray thick curve is the relation between the two for  $z<1$ (Whitaker et al. 2014). A yellow line indicates SFR = 1 $M_{\solar} yr ^{-1}$.
    }
    \label{fig:example_figure}
\end{figure}

Addionally, an interesting parameter, $f_{\mu}$, is defined by the ratio of the IR emission from coldest dust ($L_{d}^{ISM}$), compared to total dust luminosity L$_{d}^{tot}$ (or L$_{IR}$), which means the fraction of cold dust contribution to the total IR luminosty. Therefore,  $f_{\mu}$ represents what amount of IR emssion is irrelevent to current star-formation activity in the galaxy, according to the modelling by \texttt{MAGPHYS}.   
  In Fig. 5, we present the relation between SFR and $M_{\ast}$ with colorbar in each panel, showing redshift (left panels) and $f_{\mu}$ (right panels), respectively.  If the contribution from coldest dust predominates (indicating IR emission from outside SF regions is mainly responsible for L$_{IR}$),  $f_{\mu}$ approaches to 1 (see colorbar in the upper right panel). Our galaxies are  slightly crowded at the center of their gathering, but rather spread out smoothly without significant sign along the main-sequence (MS) indicated by background gray lines (Elbaz et al. 2011) in upper panels.  In the specific SFR (sSFR) vs $M_{\ast}$ plane in the lower panels, they also show wide distribution compared to the sSFR -- $M_{\ast}$ relation for $z<1$, a gray curve showing sSFR goes flat for $M_{\ast} < 10^{10} M_{\solar}$ but begins to drop at higher mass range (Whitaker et al. 2014).  This shows our sampling based on the 250 $\mu$m detection with MIR PAH feature is not biased to any specific population in terms of $f_{\mu}$   in this  SFR vs $M_{\ast}$ plane, although our selection seems slightly biased towards the brighter galaxies in the redshift range  $z < 0.5$. If our sample is dominated mostly by normal MS galaxies, a significant concentration must have appeard between $z=0.0$ and $z=1.0$ lines on upper panel or around a gray curve in lower panel (Schreiber et al. 2016; Whitaker et al. 2014).  The wide spread of our sample seems to be affected by combination of types and the redshifts. Distribution of $f_{\mu}$ on the upper right panel, along the perpendicular direction to the MS line, gives some information why galaxies are normally spread on this plane.

\begin{figure}
 \includegraphics[width=0.95\columnwidth]{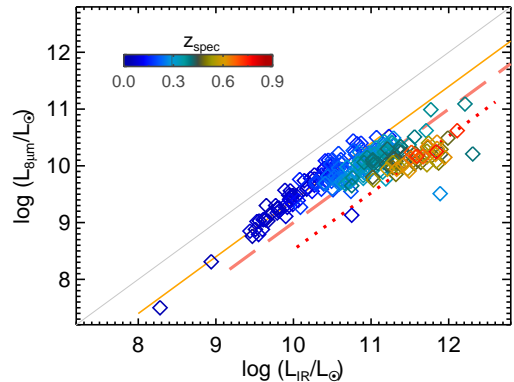}
    \caption{ Distribution of $L_{8{\mu}m}$ as a function of $L_{IR}$ with a colorbar indicating redshift. Orange-yellow line  represent $L_{IR}/L_{8{\mu}m} \sim 5$, normally known as a representative value of IR8  (Elbaz et al. 2011).  Long dashed (pink) line  represent $L_{IR}/L_{8{\mu}m} = 10$.  There are much stronger deficit in the NEP field at  $z > 0.6$ as shown by colorbar.  It is interesting that more luminous  galaxies show more severe shortage of PAH luminosity.
    } 
    \label{fig:example_figure}
\end{figure}

To reveal the physical condition of IR galaxies using their SEDs, many works (Elbaz et al. 2011; Reddy et al. 2012; Magdis et al. 2012; Murata et al. 2014; Schreiber et al; 2017)  have measured $L_{8{\mu}m}$ and compared with $L_{IR}$,  or used  IR8 (defined by $L_{IR}/L_{8{\mu}m}$), because $L_{8{\mu}m}$ dominated by the PAH features around 8 $\mu$m (e.g., at 6.2, 7.7, and 8.6 $\mu$m) is believed to give a significant clue to the interplay between the overall strength of PAH luminosity and the SF activity of a galaxy. Most of the previous works focused on the high redshift ($z>1$)  and discussed the reduced emission of PAH features and physical explanation. From the value of IR8 = 4.9 (Elbaz et al. 2011)  to the recent report of IR8 $=$ 3.5 and 7 (at $z=1$ and 2, Schreiber et al. 2018),  tried to focus on the MS galaxies, but sometimes source selection  tends to be biased to starbursts.  Most of galaxies in Elbaz et al. (2011) shows constant ratio of this value except for a very small number of sample.  It is actually difficult to track down the origin of the difference. But it might be attributed to the combination of different selection method: our sample is based on the detection of continuous MIR band, and includes various types of galaxies in terms of $f_{\mu}$  (although  biased to brighter starburst type at $z>0.5$). But we have to keep in mind that the derivation of IR8 is model dependent.      Here, in Fig. 6, we present the simple comparison of $L_{8{\mu}m}$ and $L_{IR}$, derived more accurately from   galaxies in the low-$ z (<1$) with SED analysis based on the continuously covered  MIR wavelength range as well as spectroscopic redshift. IR8 = 4.9 (Elbaz et al. 2011) track is shown in orange line. Galaxies at lowest redshift range ($z<0.3$) seem to follow this line, but beyond $z>0.4$, $L_{8{\mu}m}$ is getting weaker compared to $L_{IR}$. Long-dashed pink line and dotted red line indicates IR8 = 10 and IR = 30, respectively. As shown here,   more luminous galaxies show more severe shortage of PAH emission.   Most plausible and popular explanation is strong radiation from SF region destroys PAH molecules in the photo-dissociation regions (PDRs). AGN can not be discussed here. Another possible explanation can be applied to a few sample having large $f_{\mu}$. An unexpectedly large amount of cold dust can generate high $L_{IR}$, which can also cause relative $L_{8{\mu}m}$ shortage.  PAH deficit phenomena of our sample shows variation according to the redshift but it is not easy to assure this is evolution.    Stellar mass or metallicity also believed to be involved in some way.   Although the positive relation of metallicity to stellar mass in galaxies has been known,   the connection of PAH emission (or IR8) to metallicity seems beyond this work.  We can just make a crude guess this pattern of deficits could be related to the mass dependent evolution of galaxies (Juneau et al. 2005; Bundy et al. 2006; Marchesini et al. 2009). 


\begin{figure}
 \includegraphics[width=\columnwidth]{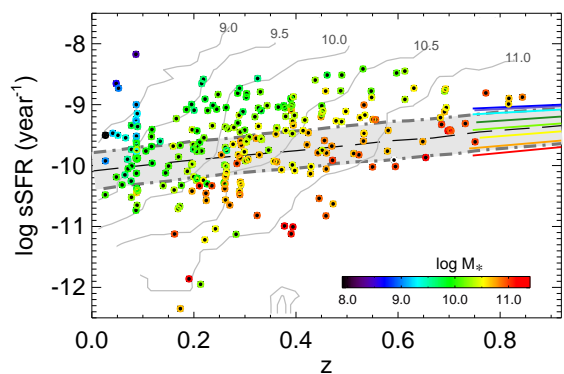}
    \caption{ Evolution of sSFR of our sample galaxies. Colorbar shows the stellar mass M$_{\ast}$ of our sample.  Gray contours across the plane indicate stellar mass whose mean values are  on the top of the figure.  Gray shaded region and color lines represent sSFR of MS galalxies (Elbaz et al. 2011; Schreiber et al. 2016). As shown here, the trend of our  galaxies shows mass-dependent time evolution clearly.
    } 
    \label{fig:example_figure}
\end{figure} 

In order to show the evolutionary properties, we present sSFR of our sample galaxies as a funciton of redshift and stellar mass (as indicated by colorbar) as shown in Fig.7.   Here, red color shows the most massive galaxies. Gray contours across the plane  represent mean stellar mass (see the numbers on the top of the figure). This plot shows most massive galaxies in a certain redshift range  have the lowest sSFR, suggesting  that more massive galaxies have relatively less amount of gas compared to the less massive galaxies,  because massive galaxies have already used up large fraction of available gas.  They have made stars more actively and rapidly, so that violently consumed up the gas in earlier epoch.  We can see this tendency continues consistently from 0.9 to current epoch, which is observational evidence of down-sizing trend (Cowie et al. 1996; Fontanot et al. 2009).  Gray shaded region indicates the redshift evolution of the median sSFR of MS galaxies from Elbaz et al. (2011).  Colored lines overlapped from $z>0.7$ are sSFR evolution of MS galaxies for different stallar mass bins (Whitaker et al. 2014).  
Because our sample includes various types from quiescent (low SFR) to starburst (high SFR), our sSFR ranges much wider than theirs. However, this trend of our sample galaxies clearly shows mass-dependent time evolution.

\section{Summary and Conclusion}
 
 Taking advantage of continuous AKARI/MIR band coverage with accurate spectroscopic redshifts, we successfully derived 8$\mu$m luminosity (dominated by PAH) of various galaxies ($z<1$) in the NEP-Wide field.  Thanks to the 250$\mu$m detection by SPIRE, total infrared luminosity was derived using reliable constraint on the FIR peak of galaxy SED.  The modeling of SED was carried out based on the energy balance argument by \texttt{MAGPHYS}, and we obtained splendid and physically meaningful fits to the observed photometric data points from optical/uv to sub-mm bands.   Our  derivation of $L_{8{\mu}m}$ is based on the SED-fit, thus, could be dependent on the templates include/generated by the software.   A few per cent of our sample seem to possibly be  composite types that might have AGN contribution more or less, but they do not affect on our main results. The sample galaxies analysed in this study have PAH features in the mid-IR,  and appear to be various types,  normally classified as SF/spiral, and ranges from quiescent to starburst in terms of SFR-$M_{\ast}$ relation. Because of the spread distribution in SFR-$M_{\ast}$ plane, it is not easy to get the clear and meaningful census on and off the main-sequence, but we can test evolutionary properties of galaxies having MIR PAHs.  Interestingly, there seem to be IR luminous population  which has abundant cold dust irrelevant to current SF activity, according to the \texttt{MAGPHYS} modelling with optical up to the sub-mm band photometries.   The shortage of PAH luminosity compared to the $L_{IR}$ seems mostly due to strong radiatioin field from SF region. Some of our galaxies in the low-z ($<1$) show unexpectedly and surprisingly severe deficit, corresponing IR8 value up to $\sim$ 30, although the results and interpretation could be model dependent.  Using sSFR vs redshift, we can see clear signal of mass dependent reshift evolution of our galaxies.

 Focusing on the exact measurement based on the spectroscopic redshifts sometimes brings difficulties with small number statistics. But recently obtained high quality HSC data on this field, which is about to release, can improve photometric redshifts of IR galaxies, including previously unmatched NEPW sources, and will give more larger sample to test these galaxy properties.  SCUBA-2 survey recently kicked off on the AKARI/NEPW field will provide more SMM sample with new information on galaxy SEDs.  Besides  accurate spectroscopic redshifts sample used in this work, extended/subsequent studies on the PAH properties using more larger galaxy sample and more strong constraint on the FIR/SMM range by upcomming data will bring more reliable physical parameters that facilitate better/detailed understanding of the IR to SMM properties for the local SF activities.

\begin{ack}
We thank the refrees for the careful reading and constructive suggestions  to improve this paper. This work is based on observations with AKARI, a JAXA project with the participation of ESA, universities, and companies in Japan, Korea, the UK, and so on. TG acknowledges the support by the Ministry of Science and Technology (MoST) of Taiwan through grant 105-2112-M-007-003-MY3.  MI acknowledges the support from the National Research Foundation of Korea (NRF) grant, No. 2017R1A3A3001362, funded by the Korea government.
\end{ack}

\end{document}